\documentclass[12pt]{article}
\usepackage{graphicx}
\usepackage{amsmath}
\usepackage{amsbsy}
\usepackage{amssymb}
\usepackage{color}

\begin{document}

\title{\textsc{Classical and quantum free motions
           in the tomographic probability representation}}
\author{V.I. Man'ko$^a$ and F. Ventriglia$^b$ \\
%EndAName
{\footnotesize \textit{$^a$P.N.Lebedev Physical Institute, Leninskii
Prospect 53, Moscow 119991, Russia}}\\
{\footnotesize {(e-mail: \texttt{manko@na.infn.it})}}\\
\textit{{\footnotesize {$^b$Dipartimento di Scienze Fisiche dell' Universit%
\`{a} ``Federico II" e Sezione INFN di Napoli,}}}\\
\textit{{\footnotesize {Complesso Universitario di Monte S. Angelo, via
Cintia, 80126 Naples, Italy}}}\\
{\footnotesize {(e-mail: \texttt{ventriglia@na.infn.it})}}}
\maketitle
\begin{abstract}
Based on a geometric picture, the example of free particle motion for both
classical and quantum domains is considered in the tomographic probability
representation. Wave functions and density operators as well as optical and
symplectic tomograms are obtained as solutions of kinetic classical and
quantum equations for the state tomograms. The difference of tomograms of
free particle for classical and quantum states is discussed.
\end{abstract}

\section{Introduction}

The geometrical methods to study the behaviour of classical systems \cite%
{beppe1} play an important role in better understanding the classical motion
and structure of the Hamiltonian and Lagrangian descriptions. Also the
geometrical picture of quantum mechanics \cite{beppe2} and a clarifying
geometrical sense of calculations in quantum domain shed new light onto
intuitively difficult quantum mechanical concepts of states and observables.

One of the applications of a geometrical view on the notion of quantum
states is the quantum tomography (see recent reviews \cite%
{ibort2009,ibort2011,asorey}). The quantum tomography theory appeared in the
connection with experiments in quantum optics \cite{raymer} where the
relation of optical tomograms \cite{BerBer,VogRis} to the Wigner function
given by integral Radon transform was used to measure a photon quantum state
identified with the state Wigner function. Optical tomograms as well as
symplectic tomograms are fair probability distribution functions. They were
suggested \cite{mancini96} to be considered as a primary notion of the
quantum state, since the tomograms contain complete information on quantum
states as wave functions and density operators. This\ property of the
tomograms turns out to be useful also in classical domain, where the
tomograms can be employed as an alternative to the probability density on
the phase space describing the states of classical particle in classical
statistical mechanics \cite{olgaJRLR97}.

Bearing in mind the geometrical considerations of quantum and classical
motions developed in \cite{beppe1,beppe2}, the aim of this paper is to give
a short review of results obtained in the tomographic probability
description of classical and quantum states using the example of the motion
of a one-dimensional free particle of mass one. In Sec. 2, we review free
particle motion in classical and quantum domains in the standard classical
and quantum mechanics. In Sec. 3, we discuss the tomographic-probability
representation of classical and quantum free motion and present the basic
equations. In Sec. 4, we consider the difference of classical and quantum
state tomograms. We conclude in Section 5.

\section{Free motion}

\subsection{Classical mechanics}

Let us recall the description of the free motion of a one-dimensional
particle in classical mechanics, when no fluctuations of position $q$ and
momentum $p$\ are present. The Hamiltonian reads (we take tha mass $m=1):$%
\begin{equation}
H=\frac{p^{2}}{2}.
\end{equation}%
\ The trajectories determined by this Hamiltonian are straight lines defined
by two integrals of motion%
\begin{equation}
p_{0}\left( q,p,t\right) =p\quad ;\quad q_{0}\left( q,p,t\right) =q-pt,
\label{inidata}
\end{equation}%
where the current position and momentum read : $q=q_{0}+p_{0}t,p=p_{0}.$

In case of position and momentum fluctuations, due for example to the
temperature of the environment where the particle moves, the state of the
particle is defined by the probability density $f\left( q,p,t\right) ,$
which provides for any time $t$ the probability to find the particle in the
phase space region $dqdp$ around the point $\left( q,p\right) :$%
\begin{equation}
dw\left( q,p,t\right) =f\left( q,p,t\right) dqdp.
\end{equation}%
Of course, such a probability density is nonnegative and normalized for any
time $t:$%
\begin{equation}
f\left( q,p,t\right) \geq 0\quad ;\quad \int f\left( q,p,t\right) dqdp=1.
\label{clanorm}
\end{equation}

The probability density is known to satisfy the Liouville kinetic equation,
that for any Hamiltonian with potential $V\left( q\right) $ reads as:%
\begin{equation}
\left[ \frac{\partial }{\partial t}+p\frac{\partial }{\partial q}-\frac{%
\partial V\left( q\right) }{\partial q}\frac{\partial }{\partial p}\right]
f\left( q,p,t\right) =0.  \label{liouv}
\end{equation}%
The solutions of this equation can be expressed in terms of two integrals of
motion, $q_{0}\left( q,p,t\right) $ and $p_{0}\left( q,p,t\right) ,$ which
have the physical meaning of providing the initial point of the particle
trajectory in phase space, so that:%
\begin{equation}
q_{0}\left( q,p,t=0\right) =q\quad ;\quad p_{0}\left( q,p,t=0\right) =p.
\end{equation}%
Thus we have for the solution $f\left( q,p,t\right) $ of the Cauchy problem
corresponding to those data the expression%
\begin{equation}
f\left( q,p,t\right) =f_{0}\left( q_{0}\left( q,p,t\right) ,p_{0}\left(
q,p,t\right) \right) .
\end{equation}%
Here $f_{0}\left( q,p\right) $ is an initial probability density in the
particle phase space. All the statistics of position and momentum can be
given by taking derivatives of the characteristic function, defined as%
\begin{equation}
\chi \left( k_{1},k_{2},t\right) =\left\langle \exp \left[ \mathrm{i}\left(
k_{1}q+k_{2}p\right) \right] \right\rangle _{f}=\int \exp \left[ \mathrm{i}%
\left( k_{1}q+k_{2}p\right) \right] f\left( q,p,t\right) dqdp.  \label{carcl}
\end{equation}

For a free particle we have the Liouville equation%
\begin{equation}
\left[ \frac{\partial }{\partial t}+p\frac{\partial }{\partial q}\right]
f\left( q,p,t\right) =0,
\end{equation}%
whose solution corresponding to the initial conditions of eq. (\ref{inidata}%
) reads%
\begin{equation}
f\left( q,p,t\right) =f_{0}\left( q-pt,p\right) .  \label{clafree}
\end{equation}%
The propagator providing the solution in the form%
\begin{equation}
f\left( q,p,t\right) =\int K\left( q,p,q^{\prime },p^{\prime },t\right)
f_{0}\left( q^{\prime },p^{\prime }\right) dq^{\prime }dp^{\prime }
\end{equation}%
is given by a product of Dirac's delta functions%
\begin{equation}
K\left( q,p,q^{\prime },p^{\prime },t\right) =\delta \left( q^{\prime
}-q+pt\right) \delta \left( p^{\prime }-p\right) .  \label{kercl}
\end{equation}

In the case when fluctuations are absent we may choose as initial density%
\begin{equation}
f_{0}\left( q,p\right) =\delta \left( q-\bar{q}\right) \delta \left( p-\bar{p%
}\right) ,  \label{inifree}
\end{equation}%
with $\bar{q}$ and $\bar{p}$\ initial position and momentum of the free
particle trajectory. Then eqs. $\left( \ref{kercl}\right) ,\left( \ref%
{inifree}\right) $ provide the probability density at any time $t\geq 0,$
also as a product of Dirac's delta functions%
\begin{equation}
f\left( q,p,t\right) =\delta \left( \bar{q}+pt-q\right) \delta \left( \bar{p}%
-p\right) .
\end{equation}

So, the density time evolution corresponds completely to a trajectory
determined by the classical Newton's law of motion.

\subsection{Quantum motion}

The quantum motion of a free particle in a standard formulation of quantum
mechanics may be described by the Schr\"{o}dinger evolution equation for the
wave function $\psi \left( q,t\right) $ when the environment influence is
absent.

In general, for the Hamiltonian operator (hereafter $\hbar =1):$
\begin{equation}
\hat{H}=\frac{\hat{p}^{2}}{2}+V\left( \hat{q}\right) ,
\end{equation}%
one has the equation%
\begin{equation}
\mathrm{i}\frac{\partial }{\partial t}\psi \left( q,t\right) =\left[ -\frac{1%
}{2}\frac{\partial ^{2}}{\partial q^{2}}+V\left( q\right) \right] \psi
\left( q,t\right) .
\end{equation}

In presence of an environment, the quantum state will be described by a
density operator $\hat{\rho},$ obeying the Von Neumann evolution equation,
which in the position representation reads as%
\begin{equation}
\mathrm{i}\frac{\partial }{\partial t}\rho \left( q,q^{\prime },t\right) =%
\left[ -\frac{1}{2}\left( \frac{\partial ^{2}}{\partial q^{2}}-\frac{%
\partial ^{2}}{\partial q^{\prime 2}}\right) +\left( V\left( q\right)
-V\left( q^{\prime }\right) \right) \right] \rho \left( q,q^{\prime
},t\right) .
\end{equation}%
Here $\rho \left( q,q^{\prime },t\right) =\left\langle q|\hat{\rho}\left(
t\right) |q^{\prime }\right\rangle $ is the density matrix in the position
representation.

For a free particle, the Schr\"{o}dinger evolution equation becomes%
\begin{equation}
\mathrm{i}\frac{\partial }{\partial t}\psi \left( q,t\right) =-\frac{1}{2}%
\frac{\partial ^{2}}{\partial q^{2}}\psi \left( q,t\right) ,
\end{equation}%
while the Von Neumann evolution equation reads%
\begin{equation}
\mathrm{i}\frac{\partial }{\partial t}\rho \left( q,q^{\prime },t\right) =-%
\frac{1}{2}\left( \frac{\partial ^{2}}{\partial q^{2}}-\frac{\partial ^{2}}{%
\partial q^{\prime 2}}\right) \rho \left( q,q^{\prime },t\right) .
\end{equation}

The position statistics of the quantum particle is expressed as%
\begin{equation}
\left\langle \hat{q}^{n}\left( t\right) \right\rangle _{\rho }=\mathrm{Tr}%
\left[ \hat{\rho}\left( t\right) \hat{q}^{n}\right] =\int q^{n}\rho \left(
q,q,t\right) dq.
\end{equation}

Analogously, for the momentum we have%
\begin{equation}
\left\langle \hat{p}^{n}\left( t\right) \right\rangle _{\rho }=\mathrm{Tr}%
\left[ \hat{\rho}\left( t\right) \hat{p}^{n}\right] .
\end{equation}%
Both these moments can be obtained by using the quantum characteristic
function, that is the mean value of the usual displacement operator

\begin{equation}
\chi \left( k_{1},k_{2},t\right) =\left\langle \exp \left[ \mathrm{i}\left(
k_{1}\hat{q}+k_{2}\hat{p}\right) \right] \right\rangle _{\rho }.
\label{carqu}
\end{equation}

In spite of the similarity of the above definition with the definition of
the classical case, eq. $\left( \ref{carcl}\right) ,$ the quantum function
cannot be considered a true characteristic function, corresponding to a
probability density on phase space. In fact, as the position and momentum
operators do not commute, the covariance term of position and momentum may
be given by eq. $\left( \ref{carqu}\right) $ only by taking into account the
Heisenberg commutations relations, $\left[ \hat{q},\hat{p}\right] =\mathrm{i}%
.$ Due to uncertainty relations, in the quantum domain there does not exist
any joint probability density of position and momentum.

For a free particle the Von Neumann evolution equation has the solution%
\begin{equation}
\hat{\rho}\left( t\right) =\exp \left[- \frac{\mathrm{i}}{2}\hat{p}^{2}t%
\right] \hat{\rho}\left( 0\right) \exp \left[ \frac{\mathrm{i}}{2}\hat{p}%
^{2}t\right] ,
\end{equation}%
which corresponds to the classical solution of eq. $\left( \ref{clafree}%
\right) .$

If there is no environment influence, the state can be described by a
normalized vector $\left\vert \psi \right\rangle $ in a Hilbert space. For
such a pure state, the corresponding density operator is the rank-one
projector $\left\vert \psi \right\rangle \left\langle \psi \right\vert =\hat{%
\rho}_{\psi },$ and the formulae for position and momentum statistics read%
\begin{equation}
\left\langle \hat{q}^{n}\left( t\right) \right\rangle _{\psi }=\mathrm{Tr}%
\left[ \hat{\rho}_{\psi }\left( t\right) \hat{q}^{n}\right] =\int
q^{n}\left\vert \psi \left( q,t\right) \right\vert ^{2}dq
\end{equation}%
and%
\begin{equation}
\left\langle \hat{p}^{n}\left( t\right) \right\rangle _{\psi }=\mathrm{Tr}%
\left[ \hat{\rho}_{\psi }\left( t\right) \hat{p}^{n}\right] =\int \psi
^{\ast }\left( q,t\right) \left( -\mathrm{i}\frac{\partial }{\partial q}%
\right) ^{n}\psi \left( q,t\right) dq.
\end{equation}%
These formulae, in contrast to the classical ones obtained from the
classical characteristic function of eq. $\left( \ref{carcl}\right) ,$ show
explicitly that a joint position and momentum probability density cannot
exist for a quantum particle.

The quantum propagator for a free particle gives the wave function at time $%
t>0$ in the form%
\begin{equation}
\psi \left( q,t\right) =\int G\left( q,q^{\prime },t\right) \psi \left(
q^{\prime },0\right) dq^{\prime },
\end{equation}%
where%
\begin{equation}
G\left( q,q^{\prime },t\right) =\frac{1}{\sqrt{2\pi \mathrm{i}t}}\exp \left[
\mathrm{i}\frac{\left( q-q^{\prime }\right) ^{2}}{2t}\right] .
\end{equation}%
These formulae induce the formulae for the free particle propagator for time
evolution of the density matrix in the position representation, which read
correspondingly:%
\begin{equation}
\rho \left( q,q^{\prime },t\right) =\int K\left( q,q^{\prime
},q_{0},q_{0}^{\prime },t\right) \rho \left( q_{0},q_{0}^{\prime },0\right)
dq_{0}dq_{0}^{\prime },
\end{equation}%
with%
\begin{equation}
K\left( q,q^{\prime },q_{0},q_{0}^{\prime },t\right) =\frac{1}{2\pi t}\exp %
\left[ \mathrm{i}\frac{\left( q-q_{0}\right) ^{2}}{2t}-\mathrm{i}\frac{%
\left( q^{\prime }-q_{0}^{\prime }\right) ^{2}}{2t}\right] .
\end{equation}%
The above formula corresponds to the classical evolution of a probability
density on phase space given by eq. $\left( \ref{kercl}\right) .$ But, in
contrast to this real classical phase space density, the above complex
quantum propagator connects density states on the configuration space.

Moreover, in the quantum domain, the classical phase space eq. $\left( \ref%
{clanorm}\right) $ is substituted by the nonnegativity and normalization
conditions for the density operator:%
\begin{equation}
\hat{\rho}\left( t\right) \geq 0\quad ,\quad \mathrm{Tr}\left[ \hat{\rho}%
\left( t\right) \right] =1.
\end{equation}

Nevertheless, the recently developed tomographic scheme \cite{ibort2009}
provides a unified description of both classical and quantum mechanics in
terms of fair probability distributions of a random variable. The
tomographic description is presented in the next section.

\section{The tomographic representation}

\subsection{Classical domain tomography}

The tomographic representation of the states of a classical particle is
introduced via Radon integral transform of the probability density $f\left(
q,p,t\right) $ on the particle phase space, i.e.%
\begin{equation}
W\left( X,\mu ,\nu ,t\right) =\int f\left( q,p,t\right) \delta \left( X-\mu
q-\nu p\right) dqdp  \label{tomcl}
\end{equation}%
The recent development which provides the modification of Radon transform by
using a window function instead of Dirac delta function is presented in \cite%
{Asorey1}.

The tomogram $W\left( X,\mu ,\nu ,t\right) $ depends on the random variable $%
X$ and two real parameters $\mu ,\nu .$ The random variable is a linear
combination of position and momentum:%
\begin{equation}
X=\mu q+\nu p  \label{xcla}
\end{equation}%
with%
\begin{equation}
\mu =s\cos \theta \quad ,\quad \nu =s^{-1}\sin \theta .
\end{equation}%
It may be interpreted as the particle position in a new phase space
reference frame, with $\theta -$rotated and $s-$scaled $q-$axis. The
tomogram defined by formula $\left( \ref{tomcl}\right) $ is nothing but a
marginal probability density done along the straight line given by eq. $%
\left( \ref{xcla}\right) .$ Thus,%
\begin{equation}
W\left( X,\mu ,\nu ,t\right) \geq 0\quad ,\quad \int W\left( X,\mu ,\nu
,t\right) dX=1.
\end{equation}%
The previous Radon transform may be inverted:%
\begin{equation}
f\left( q,p,t\right) =\frac{1}{4\pi ^{2}}\int W\left( X,\mu ,\nu ,t\right)
\exp \left[ \mathrm{i}\left( X-\mu q-\nu p\right) \right] dXd\mu d\nu ,
\label{invradcl}
\end{equation}%
so that probability densities on phase space and tomographic probability
densities can be put into one-to-one correspondence. In other words, the
introduced classical tomographic representation is equivalent to the
classical phase space representation.

The evolution equation for the classical tomogram is obtained by Radon
transforming the Liouville equation $\left( \ref{liouv}\right) $ and reads:%
\begin{equation}
\left[ \frac{\partial }{\partial t}-\mu \frac{\partial }{\partial \nu }%
-\left. \frac{\partial V\left( q\right) }{\partial q}\right\vert
_{q\rightarrow -\frac{\partial }{\partial \mu }\left( \frac{\partial }{%
\partial X}\right) ^{-1}}\nu \frac{\partial }{\partial X}\right] W\left(
X,\mu ,\nu ,t\right) =0.  \label{tomLiou}
\end{equation}

Please note that, in the force term $-\partial V\left( q\right) /\partial q,$
the argument $q$ is replaced by the operator $-\frac{\partial }{\partial \mu
}\left( \frac{\partial }{\partial X}\right) ^{-1}.$ Explicitly, the operator
$\left( \frac{\partial }{\partial X}\right) ^{-1}$ is defined in terms of a
Fourier transform as%
\begin{equation}
\left( \frac{\partial }{\partial X}\right) ^{-1}\int f\left( k\right) \exp
\left( \mathrm{i}kX\right) dk=\int \frac{f\left( k\right) }{\mathrm{i}k}\exp
\left( \mathrm{i}kX\right) dk.
\end{equation}%
Due to the presence of such a term for a generic potential, the evolution
tomographic equation is integro-differential.

In view of the previous definition of the random variable, eq. $\left( \ref%
{xcla}\right) ,$ the tomograms $\ W\left( X,1,0,t\right) $ and \ $W\left(
X,0,1,t\right) $\ are the marginal probability distributions of position and
momentum, respectively. So, the statistics may be written in unified way as%
\begin{equation}
\left\langle X^{n}\right\rangle _{\mu ,\nu ,t}=\int X^{n}W\left( X,\mu ,\nu
,t\right) dX  \label{tomstat}
\end{equation}%
This statistics may be obtained also by using the tomographic characteristic
function%
\begin{equation}
\chi \left( k,\mu ,\nu ,t\right) =\left\langle \exp \left( \mathrm{i}%
kX\right) \right\rangle _{\mu ,\nu ,t}=\int \exp \left( \mathrm{i}kX\right)
W\left( X,\mu ,\nu ,t\right) dk.
\end{equation}

The tomogram of a free particle described by the phase space density of eq. $%
\left( \ref{clafree}\right) $ is:%
\begin{equation}
W\left( X,\mu ,\nu ,t\right) =W_{0}\left( X,\mu ,\nu ,t\right) =\int
f_{0}\left( q-pt,p\right) \delta \left( X-\mu q-\nu p\right) dqdp.
\label{tompropeq}
\end{equation}%
In the propagating form of this equation,%
\[
W\left( X,\mu ,\nu ,t\right) =\int K\left( X,\mu ,\nu ,X^{\prime },\mu
^{\prime },\nu ^{\prime },t\right) W_{0}\left( X^{\prime },\mu ^{\prime
},\nu ^{\prime },0\right) dX^{\prime }d\mu ^{\prime }d\nu ^{\prime },
\]%
the tomographic classical free motion propagator reads%
\begin{equation}
K\left( X,\mu ,\nu ,X^{\prime },\mu ^{\prime },\nu ^{\prime },t\right)
=\delta \left( X-X^{\prime }\right) \delta \left( \mu -\mu ^{\prime }\right)
\delta \left( \nu +\mu t-\nu ^{\prime }\right) .  \label{tomprop}
\end{equation}%
Note that this propagator coincides with the one calculated by inserting eq.
$\left( \ref{invradcl}\right) $ into eq. $\left( \ref{tompropeq}\right) ,$
on the tomographic functions which are homogeneous of degree $-1.$

\subsection{Quantum tomography}

We preliminarily observe that the classical tomogram defined above may be
written as a mean value:%
\begin{eqnarray}
W\left( X,\mu ,\nu ,t\right) &=&\int f\left( q,p,t\right) \exp \left[
\mathrm{i}k\left( X-\mu q-\nu p\right) \right] \frac{dk}{2\pi }dqdp
\nonumber \\
&=&\left\langle \int \exp \left[ \mathrm{i}k\left( X-\mu q-\nu p\right) %
\right] \frac{dk}{2\pi }\right\rangle _{f}.
\end{eqnarray}%
Then, the quantum tomogram may be introduced as operator valued version of
the previous equation, i.e.%
\begin{eqnarray}
W\left( X,\mu ,\nu ,t\right) &=&\left\langle \int \exp \left[ \mathrm{i}%
k\left( X-\mu \hat{q}-\nu \hat{p}\right) \right] \frac{dk}{2\pi }%
\right\rangle _{\rho }  \nonumber \\
&=&\mathrm{Tr}\left[ \hat{\rho}\left( t\right) \int \exp \left[ \mathrm{i}%
k\left( X-\mu \hat{q}-\nu \hat{p}\right) \right] \frac{dk}{2\pi }\right] .
\end{eqnarray}%
In other words, a quantum tomogram may be thought of as the quantum Radon
transform of a quantum density state.

We remark that the importance of the quantum tomographic picture relies on
the possibility to measure in quantum optical experiments \cite{raymer} the
above tomographic probability density.

The quantum tomogram is again the marginal probability density of the random
variable $X$, corresponding to the spectral variable of the quantum
Hermitian operator%
\begin{equation}
\hat{X}_{\mu ,\nu }=\mu \hat{q}+\nu \hat{p}.
\end{equation}%
Again, the tomograms $\ W\left( X,1,0,t\right) $ and \ $W\left(
X,0,1,t\right) $\ are the marginal probability distributions of position and
momentum of the quantum particle, respectively. By the way, we note that the
Robertson-Schr\"{o}dinger uncertainty relations \cite{SCHR,ROB} can be
expressed in tomographic form \cite{ADVLETT,PORZIO}.

Notably, also the quantum version of the Radon transform can be inverted
\cite{ibort2009}\ for reconstructing the density state%
\begin{equation}
\hat{\rho}\left( t\right) =\frac{1}{2\pi }\int W\left( X,\mu ,\nu ,t\right)
\exp \left[ \mathrm{i}\left( X-\mu \hat{q}-\nu \hat{p}\right) \right] dXd\mu
d\nu .  \label{TomRec}
\end{equation}

In general, the statistics of the random variable $X$ is given by the same
classical formula $\left( \ref{tomstat}\right) $ by using the quntum
tomographic density.

The evolution equation of a quantum tomogram is obtained by a quantum Radon
transform of the Von Neumann equation for the density operator $\hat{\rho}%
\left( t\right) .$It reads%
\begin{equation}
\left[ \frac{\partial }{\partial t}-\mu \frac{\partial }{\partial \nu }+%
\mathrm{i}\left( \left. V\left( q\right) \right\vert _{q\rightarrow -\frac{%
\partial }{\partial \mu }\left( \frac{\partial }{\partial X}\right) ^{-1}+%
\frac{\mathrm{i}}{2}\nu \frac{\partial }{\partial X}}\right) -\mathrm{c.c.}%
\right] W\left( X,\mu ,\nu ,t\right) =0.  \label{tomvneu}
\end{equation}%
It is worthy to note that the first approximation in the parameter $\nu $
decomposition of potential $V$\ yields the tomographic version of the
classical Liouville equation $\left( \ref{tomLiou}\right) .$

For a free quantum particle $V\left( q\right) =0$ and the above equation
reduces to the classical one, which reads%
\begin{equation}
\left[ \frac{\partial }{\partial t}-\mu \frac{\partial }{\partial \nu }%
\right] W\left( X,\mu ,\nu ,t\right) =0.
\end{equation}%
Thus, the solution to the quantum tomographic evolution equation is given by
the same classical propagator $\left( \ref{tomprop}\right) .$

\section{Comparison of quantum and classical tomograms}

As we have shown, the classical and quantum states of a free particle are
described by a fair tomographic probability density of a random variable $X.$
This probability density $W\left( X,\mu ,\nu ,t\right) $ is a nonnegative
normalized function in in both classical and quantum domains. There is
another common property. Due to the homogeneity property of the Dirac's
delta function $\delta \left( \lambda x\right) =\left\vert \lambda
\right\vert ^{-1}\delta \left( x\right) ,$ the tomogram satisfies the same
homogeneity condition%
\begin{equation}
W\left( \lambda X,\lambda \mu ,\lambda \nu ,t\right) =\left\vert \lambda
\right\vert ^{-1}W\left( X,\mu ,\nu ,t\right) .
\end{equation}%
This condition reflects the scaling property of the tomograms, which can be
expressed in the form of a differential equation for the tomogram:%
\begin{equation}
\left[ X\frac{\partial }{\partial X}+\mu \frac{\partial }{\partial \mu }+\nu
\frac{\partial }{\partial \nu }+1\right] W\left( X,\mu ,\nu ,t\right) =0.
\end{equation}%
Thus, all the quantum and classical tomograms, given by their own evolution
equations, have to satisfy also the above constraint equation. One can
easily check that both classical Liouville and quantum Von Neumann equations
in tomographic form, eqs. $\left( \ref{tomLiou}\right) ,\left( \ref{tomvneu}%
\right) ,$ are compatible with the above constraint equation, because they
are scaling invariant, i.e., homogeneous of degree zero. For a free particle
this is even more apparent because the quantum and classical evolution
equations coincide.

Thus, we have to clarify differences between classical and quantum
tomograms. In general, from a group theoretical point of view, these
differences appear for instance in the inverse Radon transform formulae,
containing different representations of the Weyl-Heisenberg group discussed
in \cite{ALFRWIG}.

For a free particle, the tomographic propagator is the same for both
classical and quantum evolutions, so it is indifferent to the kind of
tomogram it propagates. So, the differences can be discussed only for the
initial tomograms. A geometric point of view can help for choosing as a tool
for such a discussion the Schr\"{o}dinger-Robertson uncertainty relations
\[
\sigma _{qq}\sigma _{pp}-\sigma _{qp}^{2}\geq \frac{1}{4}\quad (\hbar =1)
\]%
instead of the Heisenberg ones, because of their symplectic invariant
character \cite{george,dod}$.$ This means that the left hand side in the
above inequality is a constant of motion. For instance, we can choose as
initial state the Gaussian tomogram
\begin{equation}
W_{0}\left( X,\mu ,\nu ,t=0\right) =\frac{1}{\sqrt{2\pi \sigma _{XX}\left(
\mu ,\nu \right) }}\exp \left[ -\frac{X^{2}}{\sigma _{XX}\left( \mu ,\nu
\right) }\right] ,\quad \left( \left\langle X\right\rangle _{\mu ,\nu
}=0\right) ,
\end{equation}%
where%
\begin{equation}
\sigma _{XX}\left( \mu ,\nu \right) =\mu ^{2}\sigma _{qq}+\nu ^{2}\sigma
_{pp}+2\mu \nu \sigma _{qp}.
\end{equation}%
The initial covariance $\sigma _{qp}$ can be taken to be zero. The chosen
initial free particle tomogram yields a classical evolution if it does not
satisfy the Schr\"{o}dinger-Robertson uncertainty relations. So, if $\sigma
_{qq}\sigma _{pp}<1/4,$ the evolution will be classical and for any $%
t>0,\sigma _{qq}\sigma _{pp}-\sigma _{qp}^{2}<1/4.$ If the Schr\"{o}%
dinger-Robertson inequality is satisfied, the evolution can be either
classical or quantum. In other words, quantum and classical tomographic
domains have a nonempty intersection.

These considerations may be made more precise \cite{ALFRPOSTYPE} from a
group theoretical point of view. Here, we limit to state only that necessary
and sufficient condition for a tomographic function to be quantum is that
the associated group function on Weyl-Heisenberg group is of positive type
according to Naimark definition \cite{NAIMARK}$.$ This means that, if $\mu
,\nu ,\tau $ parametrize the Weyl-Heisenberg group unitary irreducible
representation%
\begin{equation}
U\left( \mu ,\nu ,\tau \right) =\mathrm{e}^{\mathrm{i}\tau }\exp \left[
\mathrm{i}\left( \mu \hat{q}+\nu \hat{p}\right) \right] ,
\end{equation}%
the function $\varphi _{t}$ defined on the group as%
\begin{equation}
\varphi _{t}\left( \mu ,\nu ,\tau \right) =\mathrm{Tr}\left[ \hat{\rho}%
\left( t\right) U\left( \mu ,\nu ,\tau \right) \right] =\mathrm{e}^{\mathrm{i%
}\tau }\int \exp \left( \mathrm{i}X\right) W\left( X,\mu ,\nu ,t\right) dX
\end{equation}%
has to be of positive type, for any $t\geq 0$.

We recall that a function $\phi $\ on a group $G$ is a positive type
function if the following quadratic form is positive semi-definite:%
\begin{equation}
\sum_{j.k=1}^{n}\lambda _{j}^{\ast }\lambda _{k}\phi \left(
g_{j}^{-1}g_{k}\right) \geq 0,\quad \left( j,k=1,...,n\right) ,
\end{equation}%
for any choice of $n$ complex numbers $\lambda _{k}$ and group elements $%
g_{k}\in G,$ for any $n\in \mathbb{N}.$

That tomographic condition amounts, in a different picture of quantum
mechanics, to the condition that the density state reconstructed out of a
tomogram by eq. $\left( \ref{TomRec}\right) $ is a nonnegative operator.

\section{Conclusions}

We resume the main points of our work. The example of a free particle was
used to show that both classical and quantum states can be represented as
fair tomographic probability densities. These densities give an alternative
unified state description for both probability densities on phase space in
classical domain and state vector or density operator in quantum domain. The
evolutions of classical and quantum states are given by tomographic kinetic
equations, equivalent to Liouville and to Von Neumann equations
respectively. Statistical properties of position and momentum are provided
by a unified formula using tomographic probability distribution for
classical and quantum motion. The necessary and sufficient conditions for a
tomogram to describe a quantum or classical state are recalled in group
theoretical terms. As a consequence, we get more understanding of the
geometrical meaning of the classical and quantum Radon transform.

\section*{Acknowledgements}

V. I. Man'ko thanks the organizers of \emph{FunInGeo}, for support and
hospitality.

\end{document}